\documentclass[%
 reprint,
 amsmath,amssymb,
 aps,
pra,
]{revtex4-1}

\usepackage{graphicx}% Include figure files
\usepackage{dcolumn}% Align table columns on decimal point
\usepackage{bm}% bold math
\usepackage{float}
\usepackage{physics}
\usepackage{xcolor}
\usepackage{etoolbox}
\apptocmd{\thebibliography}{\raggedright}{}{}
\usepackage{amsmath}
\usepackage[titletoc]{appendix}

\def \beq {\begin{equation}}
\def \eeq {\end{equation}}

\date{\today}

\begin{document}

\title{Entanglement instability in the interaction of two qubits with a common non-Markovian environment}% Force line breaks with \\

\author{G. Mouloudakis$^{1,2}$}
 \email{gmouloudakis@physics.uoc.gr}

\author{P. Lambropoulos$^{1,2}$}%

\affiliation{${^1}$Department of Physics, University of Crete, P.O. Box 2208, GR-71003 Heraklion, Crete, Greece
\\
${^2}$Institute of Electronic Structure and Laser, FORTH, P.O.Box 1527, GR-71110 Heraklion, Greece}

\date{\today}

\begin{abstract}

In this work we study the steady state entanglement between two qubits interacting asymetrically with a common non-Markovian environment. Depending on the initial two-qubit state, the asymmetry in the couplings between each qubit and the non-Markovian environment may lead to enhanced entanglement in the steady state of the system, measured in terms of the two-qubit concurrence. Our results indicate that, if a qubit-qubit interaction is also present, the two-qubit steady state concurrence is always favored by the symmetric or anti-symmetric coupling configuration. Although finite, the steady concurrence is predicted to be highly unstable in this regime as long as the interaction between the two qubits is larger than the couplings between each qubit and the non-Markovian reservoir.

\end{abstract}

\maketitle

\section{Introduction}
Protecting and controlling the quantum correlations generated between individual quantum systems is one of major tasks of our century, since such correlations are the basic resources used in quantum information and quantum computing theory \cite{ref1}. Among these quantum correlations, entanglement, which expresses the non-separability of the quantum state of a compound system, has attracted extensive attention over the past few years. Even if the achievements on the study of entanglement generation between individual parts of a compound system have been truly remarkable so far, the main difficulty of preserving the entanglement in realistic configurations is associated with the fact that the coupling of such systems to their environment leads to dissipation and therefore loss of the quantum correlations, most often in times much shorter than those needed for implementing quantum information tasks.

A key realization in the study of the dynamics of entanglement in open quantum systems, is that the entanglement between two qubits interacting with two independent Markovian environments may reduce to zero after a finite time, despite the fact that the single-qubit coherence is decaying asymptotically to zero. This phenomenon predicted by Yu and Eberly \cite{ref2}, remained known as "entanglement sudden death" (ESD) and has been also verified experimentally \cite{ref3,ref4}. In the zero-temperature limit, ESD has been shown to occur for certain classes of initial two-qubit states \cite{ref5}, while finite temperature models indicate that even for the rest of those classes of initial states, ESD will eventually occur after a finite time \cite{ref6}.

Since the discovery of ESD, a number of studies focused on the role of non-Markovianity \cite{ref7} of the surrounding environments on the entanglement dynamics of a system of qubits \cite{ref8,ref9,ref10,ref11,ref12,ref13,ref14,ref15,ref16,ref17}. The main result of these studies, is that non-Markovian reservoirs may lead to the effect of entanglement revivals for finite times after the occurrence of ESD, while the inclusion of finite temperature may accelerate the decay of entanglement and induce ESD even in the non-Markovian regime \cite{ref18,ref19}. At the same time, works on the dynamics of the entanglement between two qubits coupled to independent structured environments, with the case of photonic band-gap materials serving as a prototype \cite{ref20}, revealed the possibility of achieving entanglement trapping and therefore effective hindering of ESD \cite{ref21,ref22}.

The situation however may change drastically in the case where the qubits are interacting with a common reservoir \cite{ref23,ref24,ref25,ref26,ref27,ref28,ref29,ref30,ref31,ref32,ref33,ref34,ref35}. As has been shown in such a configuration, there exists a class of initial two-qubit states that eventually lead to non-zero steady state values of entanglement after a finite time that depends on the relevant parameters of the system \cite{ref36}. For the rest of the initial two-qubit states that do not lead to finite values of entanglement in the steady state of the system, effective strategies of entanglement preservation based on the Quantum Zeno effect have been proposed \cite{ref37}. From the viewpoint of entanglement, a relatively interesting scenario arises when one considers two qubits coupled asymmetrically to a common non-Markovian environment \cite{ref36,ref37,ref38,ref39,ref40}. Depending on the choice of the initial two-qubit state, this asymmetry in the coupling strengths between each individual qubit and the environment has been shown to potentially lead to larger two-qubit concurrence values beyond the maximum values obtained in the symmetric coupling case \cite{ref36,ref37,ref38}, although exact details on the conditions of observing this enhancement are still under investigation.

In this paper, we study the steady state entanglement between two qubits interacting asymmetrically with a common non-Markovian reservoir. We explore the conditions for steady state entanglement maximization, for various initial two-qubit states, by examining the behaviour of the steady state concurrence (SSC) in the qubit-reservoir couplings space. We uncover the cases in which SSC is favored by the asymmetric coupling configuration, demonstrating at the same time its great sensitivity to the value of the coupling between the two qubits. The phenomenon of entanglement instability is also predicted in the case where the latter becomes larger than the couplings between each qubit and the non-Markovian reservoir.

The article is organized as follows: In section II we provide a detailed theoretical formulation of the system in terms of the time-dependent Schrödinger equation (TDSE), in section III we present the main results of this study and discuss their implications, while in section IV we give a brief outline of our work providing also some concluding remarks on possible physical realisations of our system.

\section{Theory}

Our system consists of two identical interacting qubits coupled asymmetrically to a common non-Markovian reservoir, as depicted in Fig. 1. The Hamiltonian of the compound system (two qubits + reservoir) consists of three parts, namely the two-qubit Hamiltonian, $\mathcal{H}_S$, the Hamiltonian describing the evolution of the environment, $\mathcal{H}_E$ and the Hamiltonian of the interaction between the environment and each individual qubit, $\mathcal{H}_I$, i.e.

\beq
\mathcal{H} = \mathcal{H}_S + \mathcal{H}_E + \mathcal{H}_I
\eeq

where,

\begin{subequations}
\beq
\begin{split}
\mathcal{H}_S   =   \omega_e \sum_{i=1}^2 & \ket{e}_i  \bra{e}_i  + \omega_g \sum_{i=1}^2 \ket{g}_i \bra{g}_i \\
& + \frac{\mathcal{J}}{2} \left( \sigma_1^{+} \sigma_2^{-} + \sigma_1^{-} \sigma_2^{+} \right)
\end{split}
\eeq

\beq
\mathcal{H}_E = \sum_\lambda \omega_{\lambda} a_{\lambda}^{\dagger} a_{\lambda}
\eeq

\beq
\mathcal{H}_I = \sum_{i=1}^2 \sum_\lambda g_i ( \omega_{\lambda} ) \left( a_{\lambda} \sigma_i^{+} +a_{\lambda}^{\dagger} \sigma_i^{-} \right)
\eeq
\end{subequations}

In the above set of equations, $\omega_g$ and $\omega_e$ are the energies ($\hbar=1$) of the ground and the excited state of the identical qubits, respectively, $\mathcal{J}$ is the strength of the qubit-qubit interaction and $g_i ( \omega_{\lambda} )$, $i= 1, 2$, the qubit-environment coupling strength. The creation and annihilation operators of the $\lambda$-mode of the bosonic environment are denoted by $a_{\lambda}^{\dagger}$ and $a_{\lambda}$, respectively, while the qubit raising and lowering operators are given by $\sigma_i^{+}= \ket{e}_i \bra{g}_i$ and $\sigma_i^{-}= \ket{g}_i \bra{e}_i$, $i=1,2$.

In the single-excitation case, at any time $t>0$, the wavefunction of the compound system can be expressed as a linear superposition of the vectors:

\begin{figure}[t] 
	\centering
	\includegraphics[width=5cm]{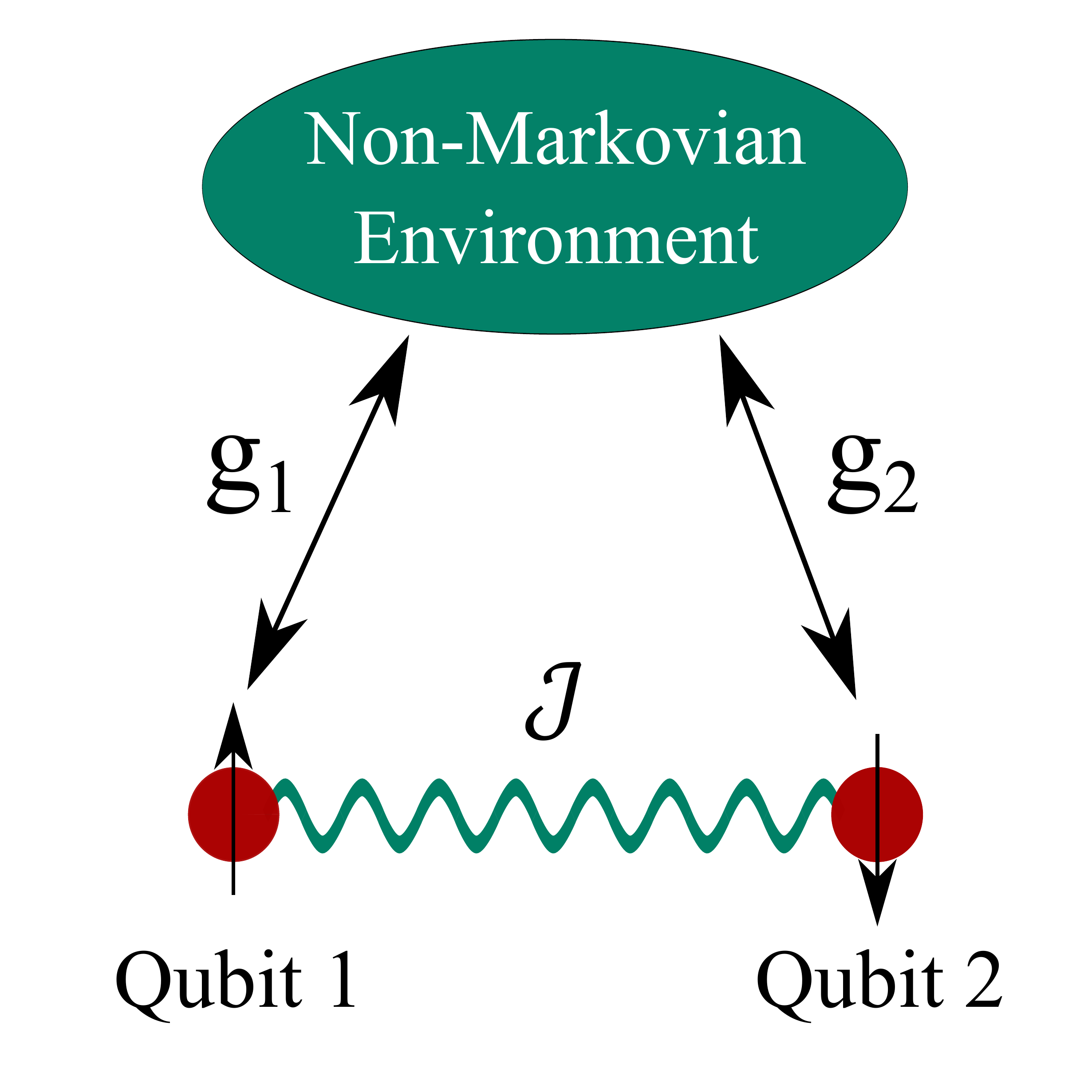}
		\caption[Fig.1]{Schematic presentation of the system at study. Two identical interacting qubits are coupled to a common non-Markovian environment with different coupling strengths.}
\end{figure}

\begin{subequations}
\beq
\ket{\psi_1} \equiv  \ket{e}_1 \ket{g}_2 \ket{0}_E
\eeq
\beq
\ket{\psi_2} \equiv  \ket{g}_1 \ket{e}_2 \ket{0}_E
\eeq
\beq
\ket{\psi^{\lambda}} \equiv  \ket{g}_1 \ket{g}_2 \ket{0 0 \dots 0 1_{\lambda} 0 \dots 0 0}_E
\eeq
\end{subequations}
according to the following expression:

\beq
\ket{\Psi (t)} = c_1(t) \ket{\psi_1} + c_2(t) \ket{\psi_2} + \sum_{\lambda} c^{\lambda}(t) \ket{\psi^{\lambda}}
\eeq

By adopting the amplitude transformation $c_i (t) = e^{-i \left( \omega_e +\omega_g \right)t} \Tilde{c}_i (t)$ , $i = 1, 2$ and $ c^{\lambda}(t) = e^ {-i \left( 2\omega_g + \omega_{\lambda} \right)t} \Tilde{c}^{\lambda}(t)$, one can easily show that the equations of motion of the tilde amplitudes according to the TDSE, reduce to:

\begin{subequations}
\beq
i \frac{ d \Tilde{c}_1 (t)}{d t} = \frac{\mathcal{J}}{2} \Tilde{c}_2 (t) + \sum_{\lambda} g_1(\omega_{\lambda}) e^{-i \Delta_{\lambda} t}   \Tilde{c}^{\lambda}(t)
 \label{Amplitude1}
\eeq
\beq
i \frac{ d \Tilde{c}_2 (t)}{d t} = \frac{\mathcal{J}}{2} \Tilde{c}_1 (t) + \sum_{\lambda}  g_2(\omega_{\lambda}) e^{-i \Delta_{\lambda} t}  \Tilde{c}^{\lambda}(t)
 \label{Amplitude2}
\eeq
\beq
 i \frac{ d \Tilde{c}^{\lambda}(t)}{d t} =  g_1(\omega_{\lambda}) e^{i \Delta_{\lambda} t} \Tilde{c}_1 (t) + g_2(\omega_{\lambda}) e^{i \Delta_{\lambda} t} \Tilde{c}_2 (t)
 \label{BathEquation}
\eeq
\end{subequations}
where $\Delta_{\lambda} = \omega_{\lambda} - (\omega_e - \omega_g) \equiv \omega_{\lambda} - \omega_{eg}$.

Formal integration of Eq. (\ref{BathEquation}) under the assumption that $\Tilde{c}^{(\lambda)}(0)=0$ and substitution back to Eqns. (\ref{Amplitude1}) and (\ref{Amplitude2}), yields:

\begin{widetext}
\begin{subequations}
\beq
\frac{ d \Tilde{c}_1 (t)}{d t} = -i \frac{\mathcal{J}}{2} \Tilde{c}_2 (t) - \int_0^t \sum_{\lambda} e^{- i \Delta_{\lambda} \left(  t - t' \right)} \Big\{ \left[  g_1(\omega_{\lambda}) \right]^2   \Tilde{c}_1 (t') +  g_1(\omega_{\lambda})  g_2(\omega_{\lambda}) \Tilde{c}_2 (t') \Big\} dt'
\label{NewAmplitude1}
\eeq
\beq
\frac{ d \Tilde{c}_2 (t)}{d t} = -i \frac{\mathcal{J}}{2} \Tilde{c}_1 (t) - \int_0^t \sum_{\lambda} e^{- i \Delta_{\lambda} \left(  t - t' \right)} \Big\{ \left[  g_2(\omega_{\lambda}) \right]^2   \Tilde{c}_2 (t') +  g_1(\omega_{\lambda})  g_2(\omega_{\lambda}) \Tilde{c}_1 (t') \Big\} dt'
\label{NewAmplitude2}
\eeq
\end{subequations}
\end{widetext}

We now replace the summation over $\lambda$ by an integration according to $\sum_{\lambda} \rightarrow \int d \omega_{\lambda} \rho (\omega_{\lambda})$, where $\rho (\omega_{\lambda})$ is the density of states (DOS) of the reservoir representing the physical environment under consideration. In what follows, we explore the case of a non-Markovian  environment, with DOS  given by the Lorentzian distribution:

\beq
\rho (\omega_{\lambda}) = \frac{1}{ \pi} \frac{\frac{\gamma}{2}}{\left( \omega_{\lambda}-\omega_c \right)^2 + (\frac{\gamma}{2})^2}
\label{DOS}
\eeq

% Check this

where $\gamma$ and $\omega_c$ are the width and the peak of the distribution, respectively. The Lorentzian DOS is widely used in the context of non-Markovian environments due to its practicality in enabling analytical solutions in systems described by complex integro-differential equations, as well as its correspondence to many realistic physical environments. As an example, Eq. ($\ref{DOS}$) could correspond to a high finesse leaky cavity with a damping rate $\gamma$, whose mode with frequency $\omega_c$ is near-resonant with the qubit energy difference $\omega_{eg}$. The generalization to any type of reservoir is straightforward by choosing the relevant DOS. 

If the coupling strengths $g_i (\omega_{\lambda})$, $i=1, 2$, vary slowly as a function of $\omega_{\lambda}$ in the vicinity of $\omega_c$, then to a good approximation, they can be substituted by their value at this frequency \cite{ref20}. In that case, in order to simplify the notation, we adopt the definitions $g_i \equiv g_i (\omega_c)$, $i=1, 2$. The resulting equations are:

\begin{subequations}
\beq
\begin{split}
\frac{ d \Tilde{c}_1 (t)}{d t}  =  -i  \frac{\mathcal{J}}{2} \Tilde{c}_2 (t) & - \int_0^t  I \left( t - t' \right) \\
& \times \left[  g_1 ^2   \Tilde{c}_1 (t') +  g_1 g_2 \Tilde{c}_2 (t') \right] dt'
\end{split}
\label{NewAmplitude1}
\eeq
\beq
\begin{split}
\frac{ d \Tilde{c}_2 (t)}{d t} = -i \frac{\mathcal{J}}{2} \Tilde{c}_1 (t) & - \int_0^t I \left( t - t' \right) \\
& \times \left[   g_2^2   \Tilde{c}_2 (t') +  g_1  g_2 \Tilde{c}_1 (t') \right] dt'
\end{split}
\label{NewAmplitude2}
\eeq
\end{subequations}

where 
\beq
\begin{split}
I \left( t-t' \right) & =  \int_{-\infty}^{+\infty} \rho (\omega_{\lambda}) e^{- i \Delta_{\lambda} \left(  t - t' \right)} d \omega_{\lambda} \\
& = \frac{\gamma}{2 \pi} \int_{-\infty}^{+\infty} \frac{e^{- i \left( \omega_{\lambda} - \omega_{eg} \right) \left(  t - t' \right)}}{\left( \omega_{\lambda}-\omega_c \right)^2 + \left( \frac{\gamma}{2} \right)^2} d \omega_{\lambda} \\
& = e^{- i \Delta_c \left(  t - t' \right)} e^{-\frac{\gamma}{2} \left(  t - t' \right)}
\end{split}
\label{ComplexIntegral}
\eeq
and $\Delta_c \equiv \omega_c - \omega_{eg}$. In view of Eqn. (\ref{ComplexIntegral}), taking the Laplace transform of Eqns. (\ref{NewAmplitude1}) and (\ref{NewAmplitude2}), one obtains:

\begin{subequations}
\beq
\begin{split}
s F_1 (s) =   \Tilde{c}_1 (0) -  i \frac{\mathcal{J}}{2}  & F_2 (s) -    g_1 ^2 \Lambda (s)  F_1 (s) \\
& -  g_1  g_2 \Lambda (s) F_2 (s) 
\end{split}
\label{Laplace1}
\eeq
\beq
\begin{split}
s F_2 (s) =   \Tilde{c}_2 (0) -  i \frac{\mathcal{J}}{2} & F_1  (s)  -    g_2 ^2 \Lambda (s)  F_2 (s) \\
& -  g_1  g_2 \Lambda (s) F_1 (s) 
\end{split}
\label{Laplace2}
\eeq
\end{subequations}
\\
where $F_1 (s)$ and $F_2 (s)$ are the Laplace transforms of the amplitudes $\Tilde{c}_1 (t)$ and $\Tilde{c}_2 (t)$, respectively, and $\Lambda(s)=\frac{1}{s+\frac{\gamma}{2}+i\Delta_c}$. The set of Eqns. (\ref{Laplace1}) and (\ref{Laplace2}) is now algebraic and can be easily solved for $F_1 (s)$ and $F_2 (s)$, the inverse Laplace of which give us the expressions of $\Tilde{c}_1 (t)$ and $\Tilde{c}_2 (t)$, respectively.

The quantity of interest is the two-qubit concurrence given by the expression \cite{ref41},
\beq
C(t) = 2 |c_1 (t) c_2^* (t)| = 2|\Tilde{c}_1 (t) \Tilde{c}_2^* (t)|
\eeq
and in particular its steady state value as a function of the coupling strengths $g_1$ and $g_2$. 

\section{Results $\&$ Discussion}

In Fig. 2 we plot the dynamics of the two-qubit concurrence for various initial two-qubit states in the non-interacting qubit case ($\mathcal{J}=0)$. In panel (a) the two qubits are interacting with the non-Markovian environment with coupling strengths $g_1/\gamma=g_2/\gamma=1$. In this arrangement we capture the well-studied case of the entanglement dynamics in the symmetrical coupling regime. The entanglement dynamics of the initial two-qubit states $\ket{\Psi \left( 0 \right)}_{12}=\ket{e}_1 \ket{g}_2$ and $\ket{\Psi \left( 0 \right)}_{12}=\frac{1}{\sqrt{2}} \left( \ket{e}_1 \ket{g}_2 \pm i \ket{g}_1 \ket{e}_2 \right)$ is symmetric with respect to the $C (t) = 0.5$ straight line, while this value is at the same time the two-qubit concurrence steady state value if the system is initially prepared in those states. If the two qubits are initially disentangled, as in the case of the initial state $\ket{\Psi \left( 0 \right)}_{12}=\ket{e}_1 \ket{g}_2$, their interaction with the non-Markovian reservoir generates entanglement. In the symmetric coupling regime considered, the concurrence dynamics of the initial state $\ket{\Psi \left( 0 \right)}_{12}=\ket{g}_1 \ket{e}_2$ is exactly the same as the respective dynamics of the $\ket{\Psi \left( 0 \right)}_{12}=\ket{e}_1 \ket{g}_2$ state, since the two qubits are assumed to be identical. On the other hand, the concurrence resulting from initially preparing the system to the states $\ket{\Psi \left( 0 \right)}_{12}=\frac{1}{\sqrt{2}} \left( \ket{e}_1 \ket{g}_2 + \ket{g}_1 \ket{e}_2 \right)$ and $\ket{\Psi \left( 0 \right)}_{12}=\frac{1}{\sqrt{2}} \left( \ket{e}_1 \ket{g}_2 - \ket{g}_1 \ket{e}_2 \right)$, follows a different dynamical picture, with the concurrence dropping from 1 to 0 after some finite time in the case of the former or being preserved to the value of 1 in the case of the latter. 

\begin{figure}[t] 
	\centering
	\includegraphics[width=7cm]{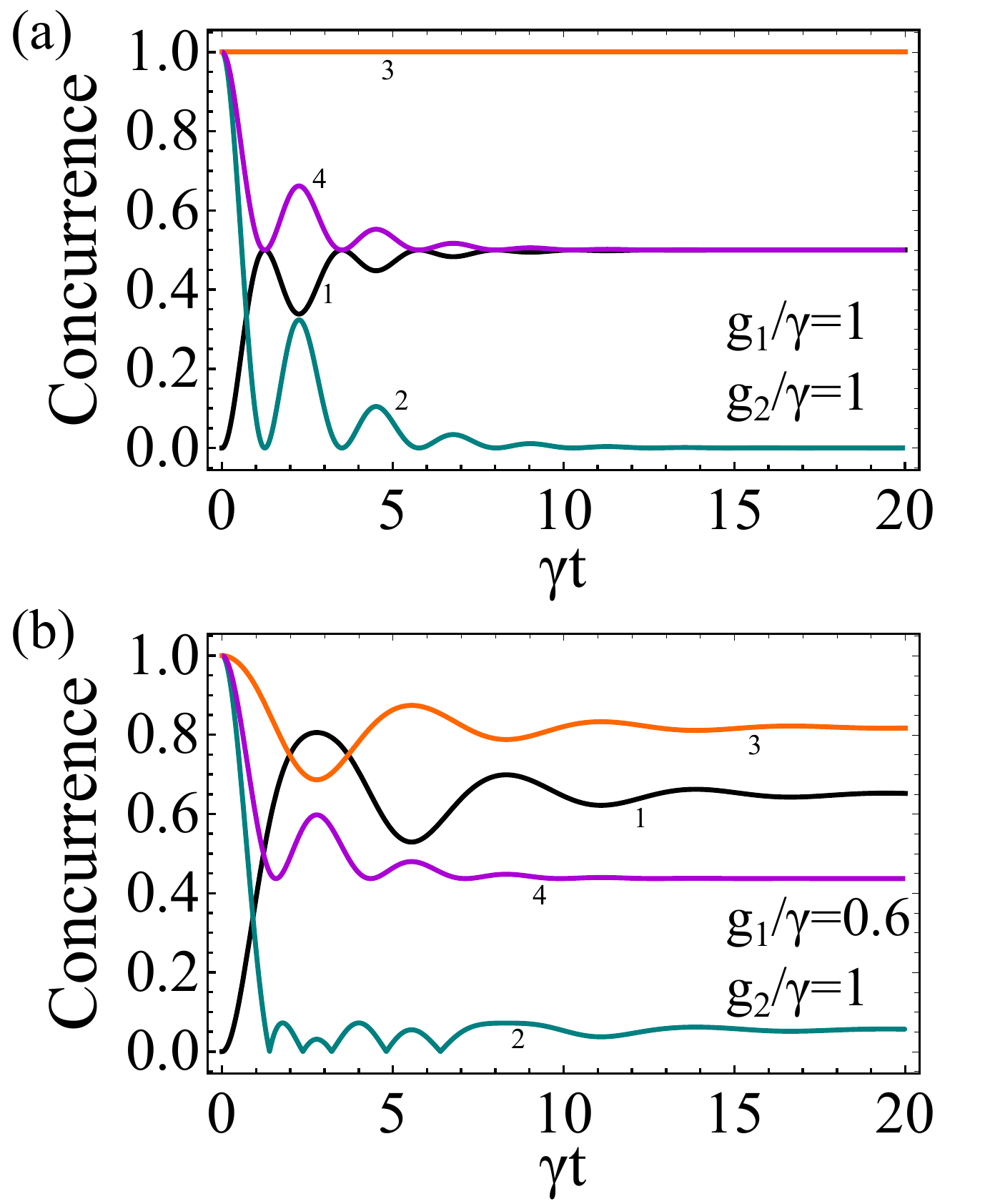}
		\caption[Fig.2]{Dynamics of the two-qubit concurrence for various initial two-qubit states. The two-qubits are assumed to be non-interacting ($\mathcal{J}=0$). (1) Black line: $\ket{\Psi \left( 0 \right)}_{12}=\ket{e}_1 \ket{g}_2$, (2) teal line: $\ket{\Psi \left( 0 \right)}_{12}=\frac{1}{\sqrt{2}} \left( \ket{e}_1 \ket{g}_2 + \ket{g}_1 \ket{e}_2 \right)$, (3) orange line: $\ket{\Psi \left( 0 \right)}_{12}=\frac{1}{\sqrt{2}} \left( \ket{e}_1 \ket{g}_2 - \ket{g}_1 \ket{e}_2 \right)$ , (4) purple line: $\ket{\Psi \left( 0 \right)}_{12}=\frac{1}{\sqrt{2}} \left( \ket{e}_1 \ket{g}_2 \pm i \ket{g}_1 \ket{e}_2 \right)$. In panel (a) the couplings of the qubits to the non-Markovian reservoir are  $g_1 / \gamma = 1$ and $g_2 / \gamma = 1$, while in panel (b) $g_1 / \gamma = 0.6$ and $g_2 / \gamma = 1$. In both panels $\Delta_c=0$.}
\end{figure}

The situation however may change drastically in the case where the two qubits interact asymmetrically with the non-Markovian environment. In Fig. (2b) we choose the values $g_1/\gamma=0.6$ and $g_2/\gamma=1$ for the qubit-environment coupling strengths. Note that, as mentioned in the caption of Fig. 2, so far we consider the case of $\Delta_c =0$, i.e. we assume that the energy difference $\omega_{eg}$ of the identical qubits is exactly on resonance with the considered Lorentzian mode of the reservoir. In this coupling configuration, all of the initial states considered above, result to totally different concurrence dynamics. The coupling asymmetry affects not only the steady state value of the two-qubit concurrence but also its characteristic oscillations that are associated with the non-Markovian character of the reservoir, and therefore the ability of information exchange between the qubits and the reservoir within some finite time that depends upon the width $\gamma$. An interesting effect of this coupling asymmetry is the possibility of concurrence stabilization to a steady state value larger than the respective steady state value of the symmetric coupling configuration \cite{ref36,ref37,ref38}, as also becomes evident by comparing the dynamics of the black line in each panel of Fig. 2.

\begin{figure}[t] 
	\centering
	\includegraphics[width=8.6cm]{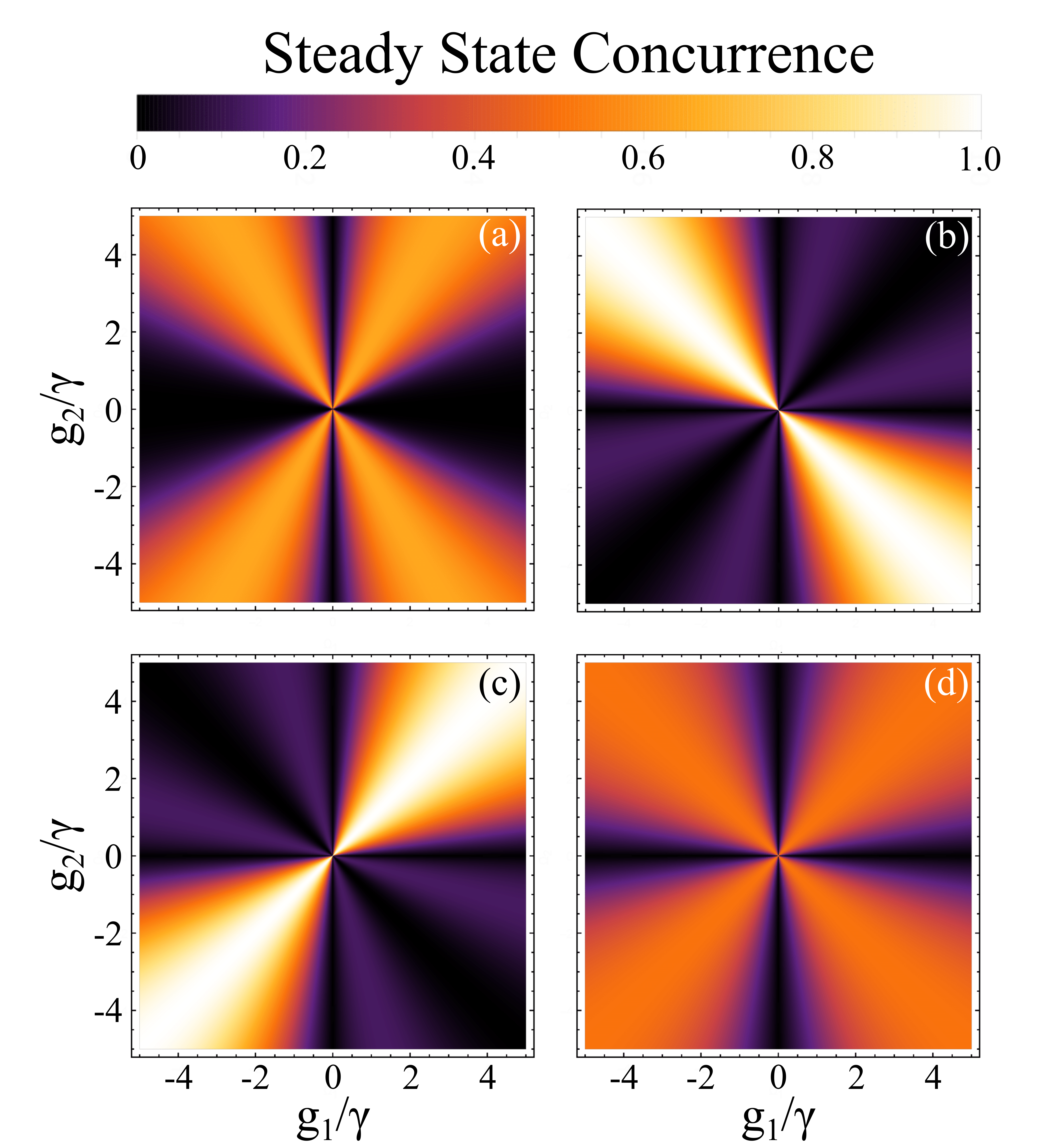}
		\caption[Fig.2]{Steady state value of the two-qubit concurrence as a function of the qubit-environment couplings $g_1$ and $g_2$, for various initial two-qubit states. The two-qubits are assumed to be non-interacting ($\mathcal{J}=0$). Panel (a): $\ket{\Psi \left( 0 \right)}_{12}=\ket{e}_1 \ket{g}_2$, panel (b): $\ket{\Psi \left( 0 \right)}_{12}=\frac{1}{\sqrt{2}} \left( \ket{e}_1 \ket{g}_2 + \ket{g}_1 \ket{e}_2 \right)$, panel (c): $\ket{\Psi \left( 0 \right)}_{12}=\frac{1}{\sqrt{2}} \left( \ket{e}_1 \ket{g}_2 - \ket{g}_1 \ket{e}_2 \right)$ , panel (d): $\ket{\Psi \left( 0 \right)}_{12}=\frac{1}{\sqrt{2}} \left( \ket{e}_1 \ket{g}_2 \pm i \ket{g}_1 \ket{e}_2 \right)$. In all panels $\Delta_c=0$.}
\end{figure}

\begin{figure}[t] 
	\centering
	\includegraphics[width=8.6cm]{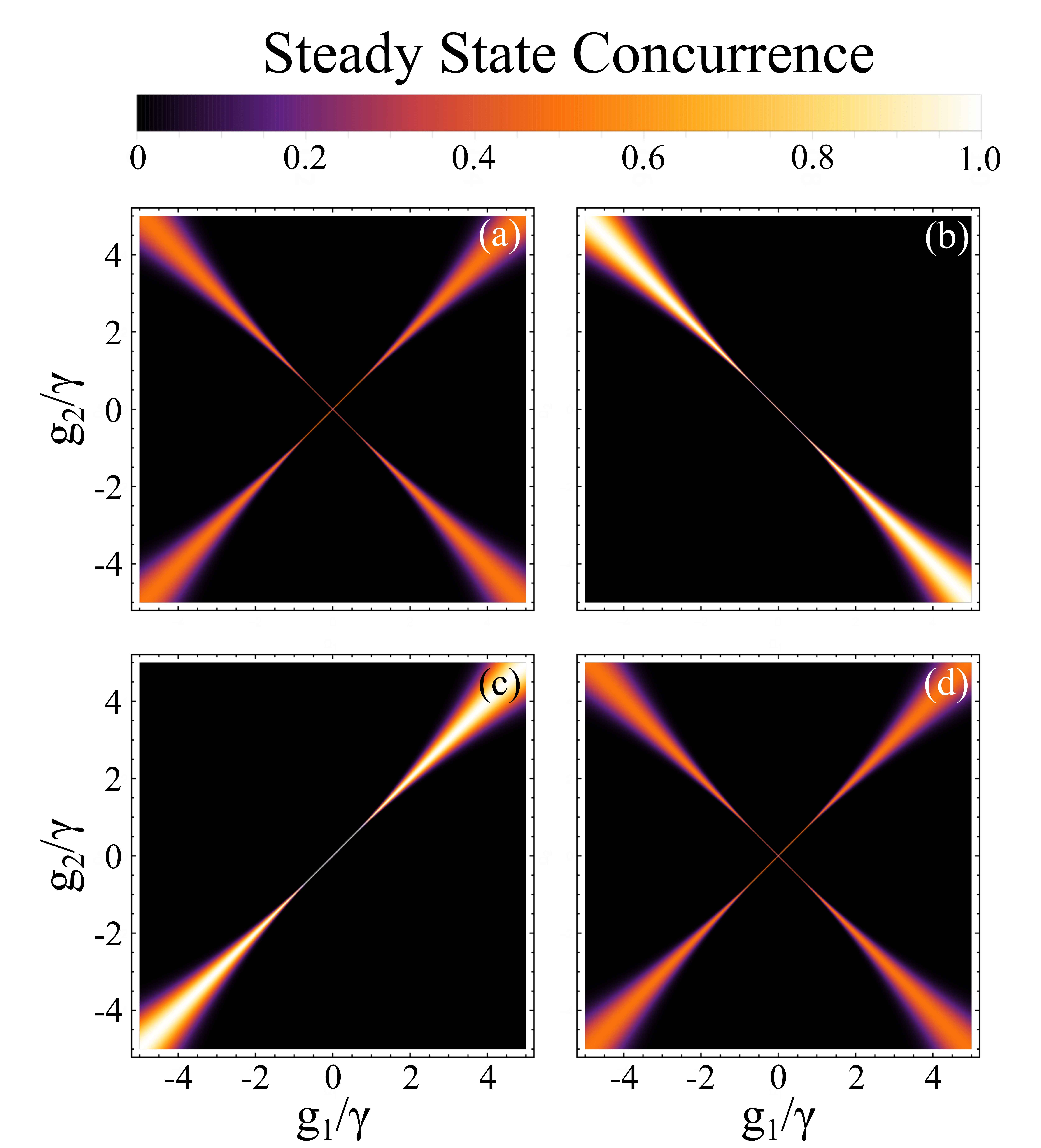}
		\caption[Fig.2]{Steady state value of the two-qubit concurrence as a function of the qubit-environment couplings $g_1$ and $g_2$, for various initial two-qubit states. The two-qubits are assumed to be interacting with a coupling strength $\mathcal{J}/ \gamma=1$. Panel (a): $\ket{\Psi \left( 0 \right)}_{12}=\ket{e}_1 \ket{g}_2$, panel (b): $\ket{\Psi \left( 0 \right)}_{12}=\frac{1}{\sqrt{2}} \left( \ket{e}_1 \ket{g}_2 + \ket{g}_1 \ket{e}_2 \right)$, panel (c): $\ket{\Psi \left( 0 \right)}_{12}=\frac{1}{\sqrt{2}} \left( \ket{e}_1 \ket{g}_2 - \ket{g}_1 \ket{e}_2 \right)$ , panel (d): $\ket{\Psi \left( 0 \right)}_{12}=\frac{1}{\sqrt{2}} \left( \ket{e}_1 \ket{g}_2 \pm i \ket{g}_1 \ket{e}_2 \right)$. In all panels $\Delta_c=0$.}
\end{figure}

\begin{figure}[t] 
	\centering
	\includegraphics[width=8.6cm]{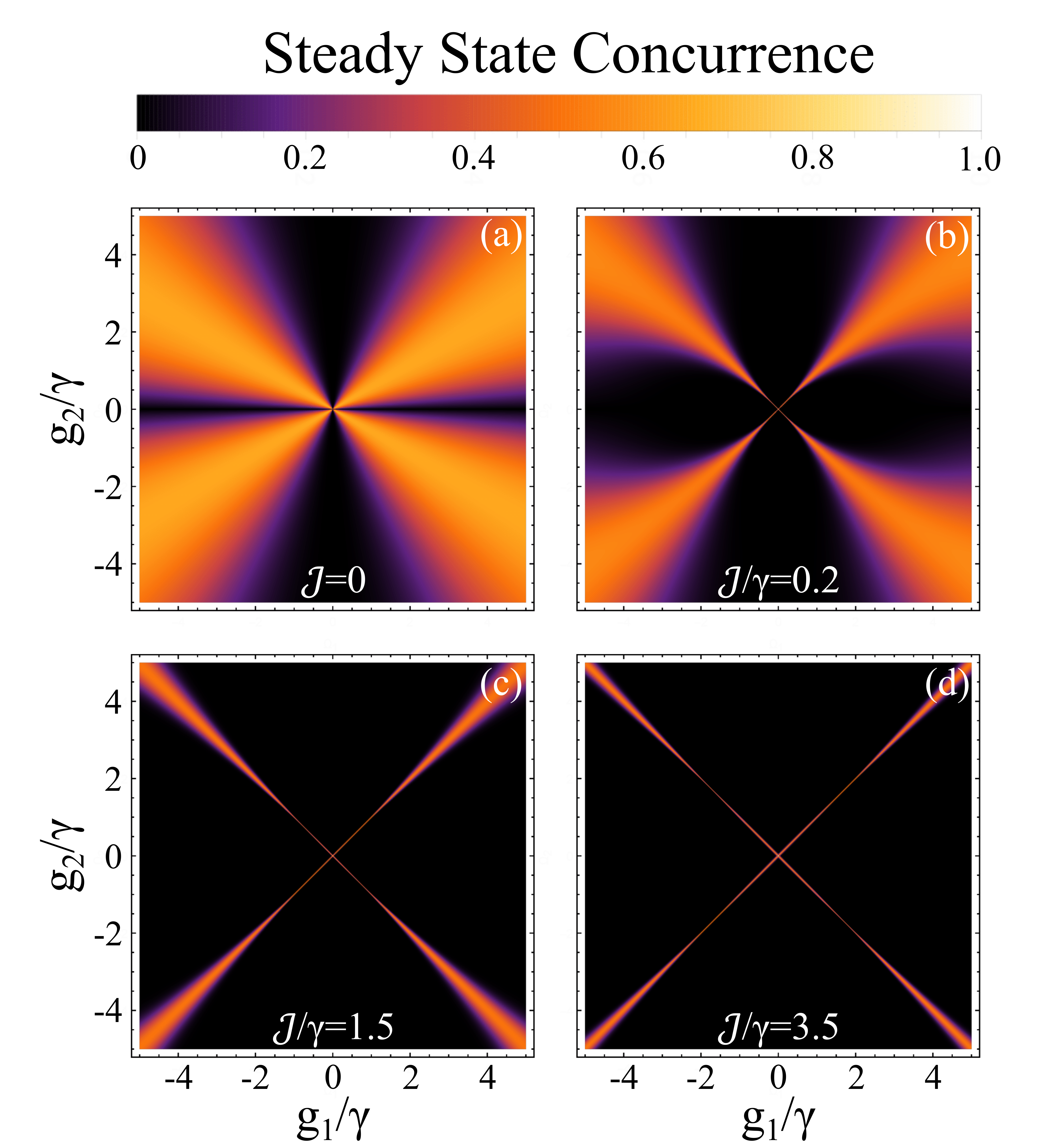}
		\caption[Fig.2]{Steady state value of the two-qubit concurrence as a function of the qubit-environment couplings $g_1$ and $g_2$ for the initial two-qubit state $\ket{\Psi \left( 0 \right)}_{12}=\ket{g}_1 \ket{e}_2$. The panels correspond to different values of the qubit-qubit coupling strength $\mathcal{J}$. Panel (a): $\mathcal{J}=0$, panel (b): $\mathcal{J}/ \gamma=0.2$, panel (c): $\mathcal{J}/ \gamma=1.5$, panel (d): $\mathcal{J}/ \gamma=3.5$. In all panels $\Delta_c=0$.}
\end{figure}

These realisations naturally lead to us to examine the profile of the SSC in the qubits' couplings space for various initial two-qubit states. The results for the non-interacting qubit case are presented in Fig. 3. Interestingly, we observe that if the two qubits are prepared initially in the state $\ket{\Psi \left( 0 \right)}_{12}=\ket{e}_1 \ket{g}_2$ (panel (a)), the SSC is not favored by the symmetric or the anti-symmetric coupling configuration, corresponding to the relations $g_1 = g_2$ and $g_1 = - g_2$, respectively. As becomes evident, there exists a big region of coupling combinations that result to SSC values larger than the value of 0.5 obtained with the symmetric and anti-symmetric configurations, going up to a maximum of 0.65. For the sake of comparison, the respective SSC pattern for the initial state $\ket{\Psi \left( 0 \right)}_{12}=\ket{g}_1 \ket{e}_2$ can be found in Fig. (5a). We should note that the reason for considering also negative values for the qubit-environment couplings is that it occurs in many physical systems that one or both couplings are negative. For example, when considering two identical qubits placed in different positions inside a cavity, if the excited mode of the cavity that is near-resonant with the qubit's transition energy is other than its fundamental, then due to the sin dependence of the coupling strength to the qubit position, there exist regions of the standing wave mode where the coupling strengths are negative. By allowing the couplings to acquire negative values as well, we came across an interesting result in the case where the initial two-qubit state is $\ket{\Psi \left( 0 \right)}_{12}=\frac{1}{\sqrt{2}} \left( \ket{e}_1 \ket{g}_2 + \ket{g}_1 \ket{e}_2 \right)$ (panel (b)). The steady state two-qubit concurrence may be zero along the symmetric couplings configuration, however it is maximum and equal to 1 along the anti-symmetric configuration and gets decreased as we move away from this region. On the other hand, if the two-qubits are prepared in the initial state $\ket{\Psi \left( 0 \right)}_{12}=\frac{1}{\sqrt{2}} \left( \ket{e}_1 \ket{g}_2 - \ket{g}_1 \ket{e}_2 \right)$ (panel (c)) the SSC pattern follows exactly the opposite picture, i.e. it is favored by the symmetric configuration, while it is zero along the anti-symmetric couplings line. The physical reason behind this effect is the following: In the absence of the interaction between the two qubits, there exists a sub-radiant two-qubit state $\ket{\Psi_{-}}$ that does not exhibit decoherence as well as a super-radiant state $\ket{\Psi_{+}}$, which is orthogonal to $\ket{\Psi_{-}}$. It is easy to show that, up to a normalization constant, these states are given by the expressions $\ket{\Psi_{-}}= g_2 \ket{e}_1 \ket{g}_2 - g_1 \ket{g}_1 \ket{e}_2$ and $\ket{\Psi_{+}}= g_1 \ket{e}_1 \ket{g}_2 + g_2 \ket{g}_1 \ket{e}_2$ \cite{ref36,ref37}. In view of the expressions above, in the symmetrical couplings configuration $g_1 = g_2$ the sub-radiant state is $\ket{\Psi_{-}} \propto \ket{e}_1 \ket{g}_2 - \ket{g}_1 \ket{e}_2$ and the super-radiant state is $\ket{\Psi_{+}} \propto \ket{e}_1 \ket{g}_2 + \ket{g}_1 \ket{e}_2$, while in the anti-symmetric couplings configuration $g_1 =- g_2$, exactly the opposite occurs. This explains the behaviour of SSC as a function of $g_1$ and $g_2$ in Figs. (3b) and (3c). 

A paradigm of a state that results to a SSC that is favored both by the symmetric and anti-symmetric couplings configuration in the same way (with a maximum value of 0.5), is the state $\ket{\Psi \left( 0 \right)}_{12}=\frac{1}{\sqrt{2}} \left( \ket{e}_1 \ket{g}_2 \pm i \ket{g}_1 \ket{e}_2 \right)$ (panel (d)). This state also indicates robustness in its resulting SSC when small deviations from the lines of these configurations are considered.

\begin{figure*}[t] 
  \includegraphics[width=16cm]{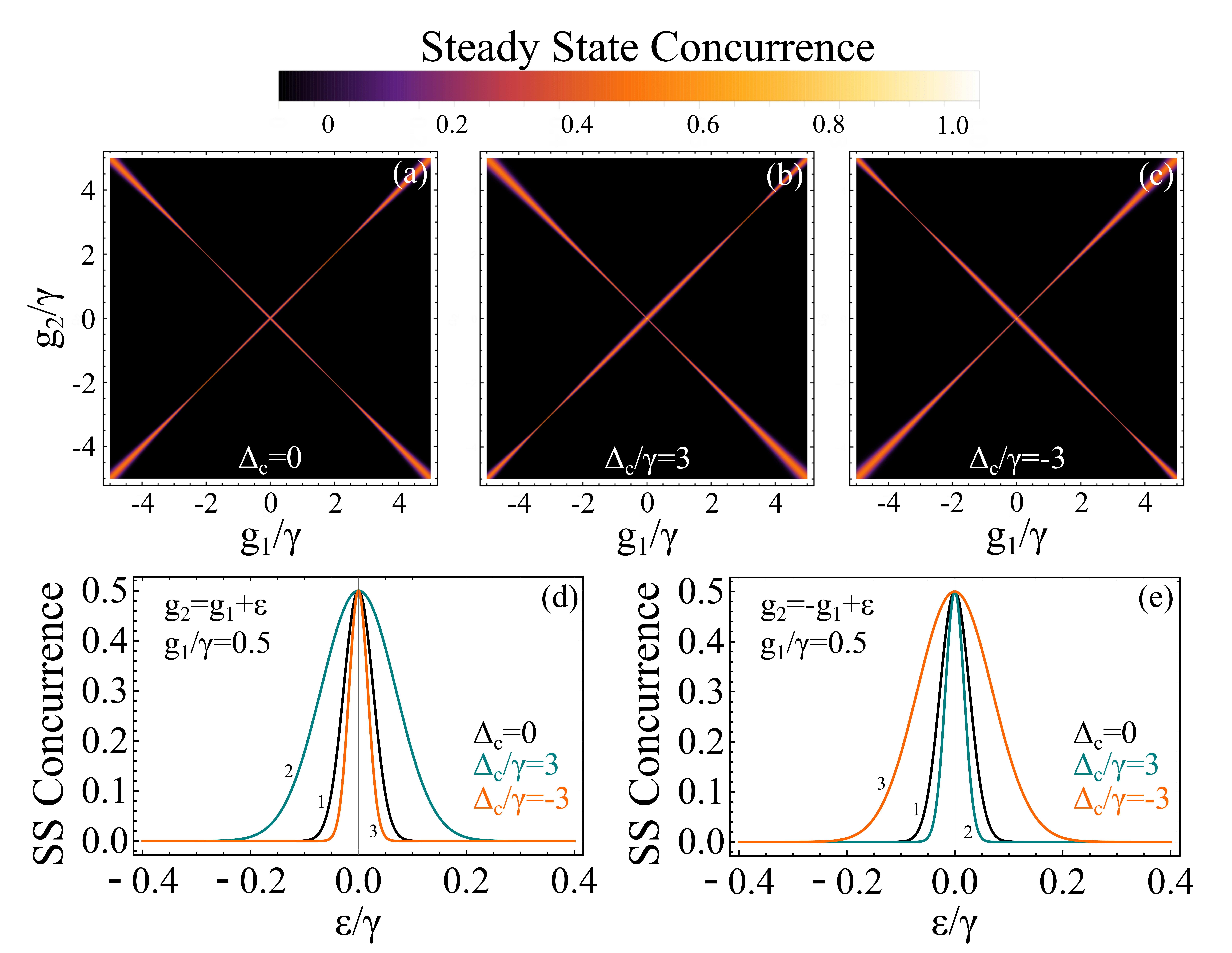}
  \caption[Fig.6]{Effects of the detuning $\Delta_c$ on the value of the two-qubit steady state concurrence. In panels (a), (b) and (c) we study the behaviour of the value of the two-qubit steady state concurrence as a function of the qubit-environment couplings $g_1$ and $g_2$, for various values of the detuning $\Delta_c$ and $\mathcal{J}/\gamma=4$. Panel (a): $\Delta_c=0$, panel (b): $\Delta_c/\gamma=3$, panel (c): $\Delta_c/\gamma=-3$. The initial state of the two-qubit system is $\ket{\Psi \left( 0 \right)}_{12}=\ket{g}_1 \ket{e}_2$. In panels (d) and (e) we show how the detuning affects the SSC when small perturbations around the symmetric and anti-symmetric coupling regimes are considered, respectively. The stability of the SSC in both cases is studied for $g_1/\gamma=0.5$. (1) Black line: $\Delta_c=0$, (2) teal line: $\Delta_c/\gamma=3$, (3) orange line: $\Delta_c/\gamma=-3$.}
\end{figure*}

The situation however changes drastically if the interaction between the two qubits is nonzero. In Fig. 4 we examine the case of $\mathcal{J}/\gamma=1$. Comparing Fig. 3 with Fig. 4, at least three major differences stand out. First, the resulting SSC pattern in the $g_1$, $g_2$ space is always favored by the symmetric or anti-symmetric couplings configuration, irrespective of the initial two-qubit state. The initial state will only determine the exact values of the SSC along the vicinity of the $g_1 = g_2$ and $g_1 = -g_2$ lines. Second, the initial states $\ket{\Psi \left( 0 \right)}_{12}=\ket{e}_1 \ket{g}_2$ (panel (4a)) and $\ket{\Psi \left( 0 \right)}_{12}=\frac{1}{\sqrt{2}} \left( \ket{e}_1 \ket{g}_2 \pm i \ket{g}_1 \ket{e}_2 \right)$ (panel (4d)) result to exactly the same SSC as a function of the qubit's positions, despite the fact that in the absence of the qubit-qubit interaction, each state had its characteristic SSC pattern. Although not presented in Fig.4, the same is true for the SSC pattern of the initial state $\ket{\Psi \left( 0 \right)}_{12}=\ket{g}_1 \ket{e}_2$. Third and most important, the SSC may be finite along the symmetric or anti-symmetric couplings configuration lines however it may exhibit high instability; namely, a slight deviation from the $g_1 = g_2$ and $g_1 = - g_2$ lines may result to zero SSC. This phenomenon of entanglement instability does not appear for any values of the qubit-qubit and qubit-environment couplings. In Fig. 5 we examine the SSC pattern of the initial state  $\ket{\Psi \left( 0 \right)}_{12}=\ket{g}_1 \ket{e}_2$ in the qubit-environment coupling space for increasing values of $\mathcal{J}$. As becomes evident, the region of entanglement instability becomes larger with the increase of $\mathcal{J}$. More precisely, the results of Fig. 5 indicate that the instability occurs for absolute values of $g_1$ and $g_2$ roughly up to the value of $\mathcal{J}$, while the regions of finite SSC along the vicinity of the symmetric and anti-symmetric couplings configurations get increasingly wider for absolute values of $g_1$ and $g_2$ larger than $\mathcal{J}$. 

These findings expand the asymptotic analysis results of Y. Li et al. \cite{ref38} who found that the only conditions for non-zero qubit's amplitudes in the steady state of the system are either the absence of the qubit-qubit coupling or the presence of equal coupling strengths between the reservoir and each qubit. While our results in Fig. 3 support the first statement, the results of Fig. 4 indicate that there exists a region of qubit-environment couplings with values slightly deviated from the vicinity of the symmetric couplings configurations that may result to finite SSC, depending on the choice of $\mathcal{J}$. We also showed that for certain initial two-qubit states this may be true only in the vicinity of the symmetric or anti-symmetric couplings lines, while for other states this can be true in both configurations (see Fig. 4). At the same time as far as the equal couplings argument is concerned, we showed that, although finite, the SSC may be highly unstable to small deviations from the $g_1 = g_2$ or $g_1 = - g_2$ lines. 

Finally, in Fig. 6 we examine the effects of a finite detuning $\Delta_c$ on the resulting SSC pattern in the $g_1$, $g_2$ space. The initial state of the two-qubit system is chosen to be $\ket{\Psi \left( 0 \right)}_{12}=\ket{g}_1 \ket{e}_2$ and the value of the qubit-qubit coupling strength is $\mathcal{J}/\gamma=4$. In Fig. (6a) the peak of the Lorentzian reservoir mode coincides with the qubit energy difference $\omega_{eg}$, i.e. $\Delta_c=0$, while in Figs. (6b) and (6c) the detunings $\Delta_c/\gamma=3$ and $\Delta_c/\gamma=-3$ are chosen, respectively. As seen in Fig. (6b), a positive detuning results to a slightly broadened region of finite SSC along the vicinity of the symmetric couplings configuration line, as long as the detuning is larger than the absolute value of the qubit-reservoir couplings. For absolute values of $g_1$, $g_2$ larger than the detuning $\Delta_c$, the region of finite SSC along the vicinity of the symmetric couplings configuration line tends to shrink. On the other hand, a negative detuning $\Delta_c$ results to exactly the same effects but along the anti-symmetric configuration line (see Fig. 6(c)). In order to probe the degree of entanglement instability for various detunings, in Figs. (6d) and (6e) we study the behaviour of the SSC when small perturbations around the symmetric and anti-symmetric coupling regimes are considered, respectively. For that reason, we choose a fixed value for one of the qubit-reservoir couplings, i.e. $g_1/\gamma=0.5$ and let the other coupling acquire the value $g_2=g_1 + \varepsilon$ (Fig. 6(d)) or $g_2=-g_1 + \varepsilon$ (Fig. 6(e)), where $\varepsilon$ is a small change in units of frequency (energy). The SSC is then studied as a function of $\varepsilon$ for various detunings. As becomes evident, a positive detuning $\Delta_c$ may increase the width of the resulting curve in the symmetric configuration but it also leads to a decreased width in the anti-symmetric configuration. At the same time, exactly the opposite picture holds on for negative detunings. Our results indicate that even if increasing the detuning can slightly improve the instability, it is ultimately a question of the precision one could achieve in the controllability and stability over the qubit-environment couplings in order to control the resulting SSC for given values of $\mathcal{J}$ and $\Delta_c$.

\section{Concluding Remarks}

In this paper we studied the steady state entanglement between two qubits that interact with a common non-Markovian environment. We showed that each initial two-qubit state results to a characteristic SSC pattern in the qubit-environment couplings space, with the maximum value of SSC occurring along the symmetric, the anti-symmetric or both coupling configuration lines. In some cases, as for example in the case where the initial two-qubit state is $\ket{\Psi \left( 0 \right)}_{12}=\ket{e}_1 \ket{g}_2$ or $\ket{\Psi \left( 0 \right)}_{12}=\ket{g}_1 \ket{e}_2$, the SSC may be favored by the choice of specific asymmetric coupling configurations. We additionally demonstrated the dramatic effects of the presence of a qubit-qubit interaction in the resulting SSC between the two qubits, with our results suggesting that in this case the SSC is finite only along the vicinity of the symmetric or anti-symmetric coupling configuration lines. It is also shown that, although finite, the SSC is predicted to be highly unstable as long as the interaction between the two qubits is larger than the couplings between each qubit and the non-Markovian reservoir; namely, a small deviation from the symmetric or anti-symmetric configuration could result to zero SSC. From the viewpoint of entanglement engineering, this result suggests that, if the exact tuning of the qubit-environment couplings to the symmetric or anti-symmetric configurations is difficult, these couplings should be larger than the qubit-qubit coupling in order to ensure the observation of finite two-qubit entanglement in the steady state of the system. The degree of instability was also studied as a function of the detuning between the peak of the considered Lorentzian mode of the reservoir and the qubits' transition energy, revealing interesting features that depend upon the parameters of the problem.

The physical realization of the considered system is within current experimental capabilities. The system can be experimentally realized using quantum dots (QDs) as qubits, since QDs can be coupled efficiently to optical cavities, \cite{ref42,ref43,ref44,ref45} while their interaction is controllable by external means, making them an ideal candidate for quantum information tasks \cite{ref46,ref47,ref48}. Alternatively, the system can be realized using trapped ions efficiently coupled to a mode of an optical resonator, as in the case of the pioneering work of B. Casabone et al. \cite{ref49}, in which the entanglement generation between the two ions was investigated. Our results acquire great significance in view of the rapid developments in the field of entanglement engineering in open quantum systems.

\section*{Acknowledgments}
GM would like to acknowledge the Institute of Electronic Structure and Laser (IESL), FORTH, for financially supporting this research through its Research and Development Program $\Pi \Delta$E00710.


\begin{thebibliography}{100}

\bibitem{ref1} See, for example, M. W. Wilde, \textit{Quantum Information Theory} (Cambridge University Press, Cambridge, 2013).

\bibitem{ref2} T. Yu and J. H. Eberly, Phys. Rev. Lett. \textbf{93}, 140404 (2004); \textbf{97}, 140403 (2006); J. H. Eberly and T. Yu, Science \textbf{316}, 555 (2007).

\bibitem{ref3} M. P. Almeida, F. de Melo, M. Hor-Meyll, A. Salles, S. P. Walborn, P. H. Souto Ribeiro and L. Davidovich, Science \textbf{316}, 579 (2007).

\bibitem{ref4} J. Laurat, K. S. Choi, H. Deng, C. W. Chou, and H. J. Kimble, Phys. Rev. Lett. \textbf{99}, 180504 (2007).

\bibitem{ref5} P. Marek, J. Lee and M. S. Kim, Phys. Rev. A \textbf{77}, 032302 (2008).

\bibitem{ref6} A. Al-Qasimi and D. F. V. James, Phys. Rev. A \textbf{77}, 012117 (2008).

\bibitem{ref7} Inés de Vega and Daniel Alonso, Rev. Mod. Phys. \textbf{89}, 015001 (2017).

\bibitem{ref8} B. Bellomo, R. Lo Franco and G. Compagno, Phys. Rev. Lett. \textbf{99}, 160502 (2007).

\bibitem{ref9} Xing Xiao, Mao-Fa Fang, Yan-Ling Li, Ke Zeng and Chao Wu, J. Phys. B: At. Mol. Opt. Phys. \textbf{42}, 235502 (2009).

\bibitem{ref10} Z. Y. Xu and M. Feng, Phys. Lett. A \textbf{373}, 1906-1910 (2009).

\bibitem{ref11} Y. J. Zhang, Z. X. Man and Y. J. Xia, Eur. Phys. J. D \textbf{55}, 173-179 (2009).

\bibitem{ref12} Qing-Jun Tong, Jun-Hong An, Hong-Gang Luo and C. H. Oh
Phys. Rev. A \textbf{81}, 052330 (2010).

\bibitem{ref13} Huang Li-Yuan and Fang Mao-Fa, Chin. Phys. B \textbf{19}, 090318 (2010).

\bibitem{ref14} B. Bellomo, G. Compagno, R. Lo Franco, A. Ridolfo and S. Savasta, Phys. Scr. \textbf{T143}, 014004 (2011).

\bibitem{ref15} Zhong-Xiao Man, Yun-Jie Xia and Nguyen Ba An, Phys. Rev. A \textbf{86}, 052322 (2012).

\bibitem{ref16} Chen Wang and Qing-Hu Chen, New Journal of Physics \textbf{15}, 103020 (2013).

\bibitem{ref17} Hong-Mei Zou, Mao-Fa Fang, Bai-Yuan Yang, You-Neng Guo, Wei He and Shi-Yang Zhang, Phys. Scr. \textbf{89}, 115101 (2014).

\bibitem{ref18} B. Bellomo, R. Lo Franco and G. Compagno, Phys. Rev. A \textbf{77}, 032342 (2008).

\bibitem{ref19} Hong-Mei Zou and Mao-Fa Fang, Chin. Phys. B \textbf{25}, 090302 (2016).

\bibitem{ref20} P. Lambropoulos, G. M. Nikolopoulos, T. R. Nielsen and S. Bay, Rep. Prog. Phys. \textbf{63}, 455 (2000).

\bibitem{ref21} B. Bellomo, R. Lo Franco, S. Maniscalco and G. Compagno, Phys. Rev. A \textbf{78}, 060302(R) (2008).

\bibitem{ref22} B. Bellomo, R. Lo Franco, S. Maniscalco and G. Compagno, Phys. Scr. \textbf{T140}, 014014 (2010).

\bibitem{ref23} D. Braun, Phys. Rev. Lett. \textbf{89}, 277901 (2002).

\bibitem{ref24} S. Oh and J. Kim, Phys. Rev. A \textbf{73}, 062306 (2006).

\bibitem{ref25} L. D. Contreras-Pulido and R. Aguado, Phys. Rev. B \textbf{77}, 155420 (2008).

\bibitem{ref26} C. Anastopoulos, S. Shresta and B. L. Hu, Quantum Inf. Process. \textbf{8}, 549 (2009).

\bibitem{ref27} K. Härkönen, F. Plastina, and S. Maniscalco, Phys. Rev. A \textbf{80}, 033841 (2009).

\bibitem{ref28} N. B. An, J. Kim and K. Kim, Phys. Rev. A \textbf{82}, 032316 (2010); Phys. Rev. A \textbf{84}, 022329 (2011).

\bibitem{ref29} X. Zhao, J. Jing, B. Corn and T. Yu, Phys. Rev. A \textbf{84}, 032101 (2011).

\bibitem{ref30} Ji Ying-Hua \textit{et al}. Chin. Phys. B \textbf{20} 070304 (2011).

\bibitem{ref31} C. H. Fleming, N. I. Cummings, C. Anastopoulos and B. L. Hu, J. Phys. A: Math. Theor. \textbf{45}, 065301 (2012).

\bibitem{ref32} J. Ma, Z. Sun, X. Wang and F. Nori, Phys. Rev. A \textbf{85}, 062323 (2012).

\bibitem{ref33} Z. X. Man, Y. J. Xia and N. B. An, Phys. Rev. A \textbf{86}, 012325 (2012).

\bibitem{ref34} L. Memarzadeh and S. Mancini, Phys. Rev. A \textbf{87}, 032303 (2013).

\bibitem{ref35} N. B. An, Phys. Lett. A \textbf{377}, 2520 (2013).

\bibitem{ref36} F. Francica, S. Maniscalco, J. Piilo, F. Plastina and K.-A. Suominen, Phys. Rev. A \textbf{79}, 032310 (2009).

\bibitem{ref37} S. Maniscalco, F. Francica, R. L. Zaffino, N. L. Gullo and F. Plastina, Phys. Rev. Lett. \textbf{100}, 090503 (2008).

\bibitem{ref38} Y. Li, J. Zhou and H. Guo, Phys. Rev. A \textbf{79}, 012309 (2009).

\bibitem{ref39} S. Golkar and M. K. Tavassoly,  Chin. Phys. B \textbf{27}, 040303 (2018).

\bibitem{ref40} L. Yan-Ling and F. Mao-Fa, Chin. Phys. B \textbf{20}, 100312 (2011).

\bibitem{ref41} W. K. Wootters, Phys. Rev. Lett. \textbf{80}, 2245 (1998).

\bibitem{ref42} K. Hennessy, A. Badolato, M. Winger, D. Gerace, M. Atatüre, S. Gulde, S. Fält, E. L. Hu and A. Imamoğlu, Nature \textbf{445}, 896–899 (2007).

\bibitem{ref43} P. Zhang, G. Song and L. Yu, Photonics Research \textbf{6}, 182-185 (2018).

\bibitem{ref44} H. Kim, D. Sridharan, T. C. Shen, G. S. Solomon and E. Waks, Optics Express \textbf{19}, 2589-2598 (2011).

\bibitem{ref45} D. Najer, et al. Nature \textbf{575}, 622-627 (2019)

\bibitem{ref46} H. R. Wei and F. G. Deng, Scientific Reports \textbf{4}, 7551 (2014).

\bibitem{ref47} X. Wang, M. Feng  and B. C. Sanders, Phys. Rev. A \textbf{67}, 022302 (2003).

\bibitem{ref48} See, for example, P. Michler, \textit{Quantum Dots for Quantum Information Technologies} (Springer, Berlin, 2017).

\bibitem{ref49} B. Casabone, A. Stute, K. Friebe, B. Brandstätter, K. Schüppert, R. Blatt and T. E. Northup, Phys. Rev. Lett. \textbf{111}, 100505 (2013)

\end{thebibliography}
\end{document}